\documentclass[twocolumn,tighten]{aastex62}

\usepackage{amsmath}
\usepackage{multirow}
\usepackage{booktabs}

\begin{document}

%\title{Resolving Protoplanetary Disk Substructures with a Future ALMA Extension}

\title{ Investigating the Future Potential of an Upgraded ALMA to Image Planet Forming Disks at Sub-au Scales}

\author{Benjamin P. Burrill}
\affiliation{Department of Physics and Astronomy, California State University Northridge, 18111 Nordhoff Street, Northridge, CA 91330, USA}

\author{Luca Ricci}
\affiliation{Department of Physics and Astronomy, California State University Northridge, 18111 Nordhoff Street, Northridge, CA 91330, USA}
%\affiliation[*]{Corresponding author.} 

\author{Sarah K. Harter}
\affiliation{Department of Physics \& Astronomy, University of Rochester, Rochester, NY 14627, USA}

\author{Shangjia Zhang}
\affiliation{Department of Physics and Astronomy, University of Nevada Las Vegas, Las Vagas, NV 89154, USA}

\author{Zhaohuan Zhu}
\affiliation{Department of Physics and Astronomy, University of Nevada Las Vegas, Las Vagas, NV 89154, USA}

%\author{Neil M. Phillips}
%\affiliation{European Southern Observatory, Karl-Schwarzschild-Strasse 2, Garching bei Muenchen, 85748, Germany}

\correspondingauthor{Luca Ricci}
\email{luca.ricci@csun.edu}

\begin{abstract}
In recent years, ALMA has been able to observe large-scale substructures within protoplanetary disks. 
Comparison with the predictions from models of planet-disk interaction has indicated that most of these disk substructures can be explained by the presence of planets with the mass of Neptune or larger at orbital radii of $\approx 5 - 100$ au. Better resolution is needed to observe structures closer to the star, where terrestrial planets are expected to form, as well as structures opened by planets with masses lower than Neptune. We investigate the capabilities of a possible extension to ALMA that would double the longest baseline lengths in the array to detect and resolve disk substructures opened by Earth-mass and Super Earth planets at orbital radii of $1-5$ au. By simulating observations of a family of disk models using this extended configuration in ALMA Bands 6 and 7, we show that an upgraded ALMA would detect gaps in disks formed by super-Earths as close as 1 au, as well as Earth-mass planets down to $2-3$ au from the young host stars in nearby star forming regions.  

%Such planets may be observable by direct imaging with future observatories, which could help validate models of planet formation.
\end{abstract}

\section{Introduction}

Planets form out of the gas and dust contained in disks orbiting young pre-Main Sequence stars \citep[see][for a recent review]{Andrews:2020}. 
Besides providing the raw material for the development of planetary cores and atmospheres, these circumstellar disks also play a key role in the dynamics of the newly born planetary systems. This is because the physical interaction between the disk and planets naturally leads to an exchange of angular momentum, with significant consequences on both the planets and the disk within the same system~\citep[][]{Kley:2012}. 

The main effect of the change in the planet angular momentum is an evolution of the planet orbit, a process known as planet \textit{migration}~\citep{Goldreich:1980,Lin:1986}. In turn, the change in the angular momentum of the disk can lead to significant perturbations into the disk physical structure. The morphology and physical characteristics of these perturbations, or \textit{substructures}, can strongly modify the subsequent evolution of the system, both by triggering or inhibiting the formation of a new generation of planets, and by altering the disk-planet interaction~\citep[][]{Baruteau:2014}.

Observationally, the direct detection and characterization of these disk substructures has therefore two main purposes. The first is that they allow us to test the predictions of physical models of the disk-planet interaction~\citep[e.g.,][]{Dong:2015b,Dong:2015a}. If the properties of a companion in a real system are known, the detailed comparison between observations and theoretical models can help us to refine our theoretical models. %Only from a detailed comparison between the predicted and observed disk substructures we can identify what is the frequency in real systems of the physical processes proposed by theories of planet formation and disk-planet interaction.
Moreover, the detection of disk substructures can be used as a way to unveil the presence of young planets in these systems. Although this is an indirect method for the detection of exoplanets, in some systems this may provide the only viable method to detect exoplanets, especially those that are highly embedded within the disk~\citep[][]{Sanchis:2020}. The detailed physical modelling of the observed substructures can then be used to extract estimates for some of the main planetary parameters, such as planet mass and orbital radius \citep[e.g.,][]{Jin:2016,Dipierro:2018}, after accounting for the known degeneracies between the properties of the disk substructures, planet mass, gas viscosity and disk aspect ratio~\citep[e.g.,][]{Fung:2014}.

One of the most spectacular successes of the \textit{Atacama Large Millimeter/submillimeter Array} (ALMA) has been the imaging at (sub-)millimeter wavelengths of several substructures in the dust continuum emission from disks located in nearby star forming regions (distances $< 200$ pc).
The results of the first ALMA surveys with angular resolution lower than about $0.1$ arcsec, corresponding to spatial resolutions $< 15-20$ au,  indicate that disk substructures are nearly ubiquitous in the mapped systems~\citep[][]{Andrews:2018,Cieza:2021}, with the majority of them showing nearly concentric annular rings and gaps~\citep[][]{Huang:2018}. 

In several cases, the morphology of these features is in line with the predictions from models of the disk-planet interaction, and estimates for the planet masses and orbital radii have been extracted from the comparison between the substructures predicted from the models and those observed with ALMA. 
For example, \citet[][]{Zhang:2018} have used the results of global 2D hydrodynamical simulations with gas and dust coupled to a radiative transfer code to obtain synthetic images for the dust continuum emission at the wavelengths of the \textit{Disk Substructures at High Angular Resolution Project} (DSHARP) ALMA Large Program. From a close comparison with the annular substructures observed in several of the DSHARP targets, they inferred relatively large planet masses ranging between the mass of Neptune up to $\sim 10~M_{\rm{Jup}}$, and orbital radii between about 10 and 100 au. Similar ranges for these planetary properties were obtained by \citet[][]{Lodato:2019}, who combined the DSHARP sample with those from other ALMA observations of nearby disks at similar wavelengths~\citep[][]{Long:2018}. 

As highlighted in these studies, the lack of planets with mass below Neptune and with orbital radii $< 10$ au, more in line with the exoplanets typically observed around Main Sequence stars, is almost certainly due to the limited angular resolution of the ALMA observations. The width of a gap opened by a planet in the disk gets smaller for planets with lower mass and orbiting closer to the star~\citep[][]{Kley:2012}. Hence, the only way to alleviate the observational limitations and probe disk substructures due to terrestrial planets is by observing disks with better resolution than it is possible with the current ALMA interferometer.

Recently, \citet[][]{Ricci:2018} and \citet[][]{Harter:2020} presented investigations on the potential of a future \textit{Next Generation Very Large Array} to detect the signatures of young planets in the dust continuum emission of disks at wavelengths of 3 mm and longer. A similar investigation has been presented by \citet[][]{Ilee:2020} on predictions of future radio observations with the \textit{Square Kilometre Array}. In this work we present the potential of a possible future extension of the current ALMA array to detect and spatially resolve disk substructures due to Earth-mass and Super Earth planets at orbital radii lower than 10 au, i.e. in the disk regions where terrestrial planets are expected to form. The extension to the ALMA array considered here was inspired by the investigations discussed in the \textit{ALMA Development Program: Roadmap to 2030} \citep[][]{Carpenter:2020}, and consists of an array configuration with longest baselines of about 32 km, a factor of $2\times$ longer than the current ALMA longest baselines. 

Section~\ref{sec:methods} describes the methods used for our investigation, Section~\ref{sec:results} presents the main results obtained in this work. After a discussion in Section~\ref{sec:discussion}, the main conclusions are summarized in Section~\ref{sec:conclusions}.

\section{Methods}
\label{sec:methods}
In this section we describe the methods used for our investigation. These comprise global 2D hydrodynamic simulations that calculate the disk-planet interaction in a disk made of gas and dust particles, the derivation of synthetic images for the dust continuum emission from these models, and finally the simulation of observations with the current ALMA as well as with an extended ALMA with longer baselines.  

\subsection{Hydrodynamic Simulations for the Disk-Planet Interaction}
\label{sec:model}

The methods used for the hydrodynamic simulations that calculate the time evolution of a disk made of gas and dust particles gravitationally interacting with a planet are the same as in \citet{Zhang:2018}
and
\citet{Harter:2020}, which are based on the Dusty FARGO-ADSG code~\citep[][]{Masset:2000,Baruteau:2008a,Baruteau:2008b,Baruteau:2016}. 

We follow the same parametrization scheme as in~\citet{Harter:2020}, with a disk with an initial gas surface density $\Sigma_{g} = \Sigma_{g}(r) = \Sigma_{g,0} (r/a_p)^{-1}$, with $r$ being the stellocentric radius, $a_p$ the planet orbital radius\footnote{Our models do not account for planet migration.}, and $\Sigma_{g,0} = \Sigma_{g} (r = a_p)$ a normalization factor. Although at the beginning of the simulations the disk is azimuthally symmetric, this symmetry is broken early on because of the interaction with the planet. Since our main goal is to test the potential of an extended ALMA to detect disk substructures due to planets in the terrestrial planet forming regions of disks, in our simulations we considered planet orbital radii $a_p = 1, 2, 3$ and 5 au. 
Our numerical grid is made of 750 grid points in the radial direction spanning the interval $[0.1a_p,10a_p]$, and 1024 grid points covering the full azimuthal direction $[0,2\pi]$. 
In order to explore the dependence of the results of our investigation on the disk surface density (hence, disk mass), we ran simulations with $\Sigma_{g,0}$ values of $100, 300, 600, 1200, 2400\,{\rm g/cm^2}$. The initial dust surface density is 1/100 of the initial gas surface density and our simulations account for a power-law grain size distribution with power-law exponent equal to $-3.5$~\citep[see e.g.,][]{Ricci:2010}, and truncated at a maximum grain size of 1 cm \citep[see][]{Zhang:2018,Harter:2020}.

The models assume a local isothermal equation of state with temperature $T$ related to the pressure scale height $h$ via the relation for vertical hydrostatic equilibrium: $h/r = c_s / v_K$, where $c_s$ is the local sound speed $c_s = \sqrt{RT/\mu}$, with an assumed mean molecular weight of the gas $\mu = 2.35$, and $v_K$ is the local Keplerian velocity. As in \citet[][]{Harter:2020}, our simulations assume a value of 0.03 for the vertical aspect ratio $h/r$ at the radial location of the planet, and a constant value of $10^{-5}$ for the  Shakura-Sunyaev $\alpha$-parameter for the gas viscosity across the disk~\citep[][]{Shakura:1973}. Our model considers a \textit{passive} disk, in which the temperature is determined by the stellar luminosity $L_{\star}$ and disk flaring factor $\phi$ via $T(r) = (\phi L_{\star}/8\pi\sigma r^2)^{1/4}$, where $\sigma$ is the Stefan-Boltzmann constant. Our models were calculated with $L_\star = 10~L_{\odot}$ and $\phi = 0.02$ \citep[][]{DAlessio:2001}. Any contribution to the disk temperature by internal viscous heating was neglected. 

In these models the planet and stellar mass, $M_p$ and $M_{\star}$, respectively, enter via the ratio $q = M_p/M_{\star}$, and for this work we launched simulations with  two different $q$ values of 1 and 10~$M_{\oplus}/M_{\odot}$, representing planets with mass of the Earth and of a Super Earth around a Solar-mass star, respectively. Our simulations were run for 7000 and 5000 planetary orbits for $q = 1$ and $10~M_{\oplus}/M_{\odot}$, respectively.

The results of these simulations were used to produce synthetic maps for the dust thermal continuum emission at two wavelengths, 0.88 and 1.25 mm, corresponding to ALMA Band 7 and 6, respectively, following the method detailed in \citet{Zhang:2018}. These wavelengths were chosen as these spectral bands provide the best sensitivity to thermal emission from dust with ALMA. To derive these maps we assumed a distance to the disk model of 140 pc, similar to the distance to nearby star forming regions which are routinely observed with ALMA.

\subsection{Simulated ALMA observations}

The synthetic maps described in the previous section for the dust continuum emission derived from our models were used to predict the results of future observations with ALMA. This was done using the version 5.7 of the \texttt{CASA} software package\footnote{\url{https://casa.nrao.edu}}
\citep{McMullin:2007}. 

For these calculations, we considered two main array configurations. To simulate observations with the longest baselines in the \textit{current} ALMA array, we
considered the array configuration file \texttt{alma.cycle6.10.cfg} available in \texttt{CASA}\footnote{\url{https://almascience.nrao.edu/tools/documents-and-tools/cycle6/alma-configuration-files}},
with baselines ranging from 256 m to 16.2 km.

\begin{figure}[hbtp]
    \begin{center}
    \includegraphics[width=\columnwidth]{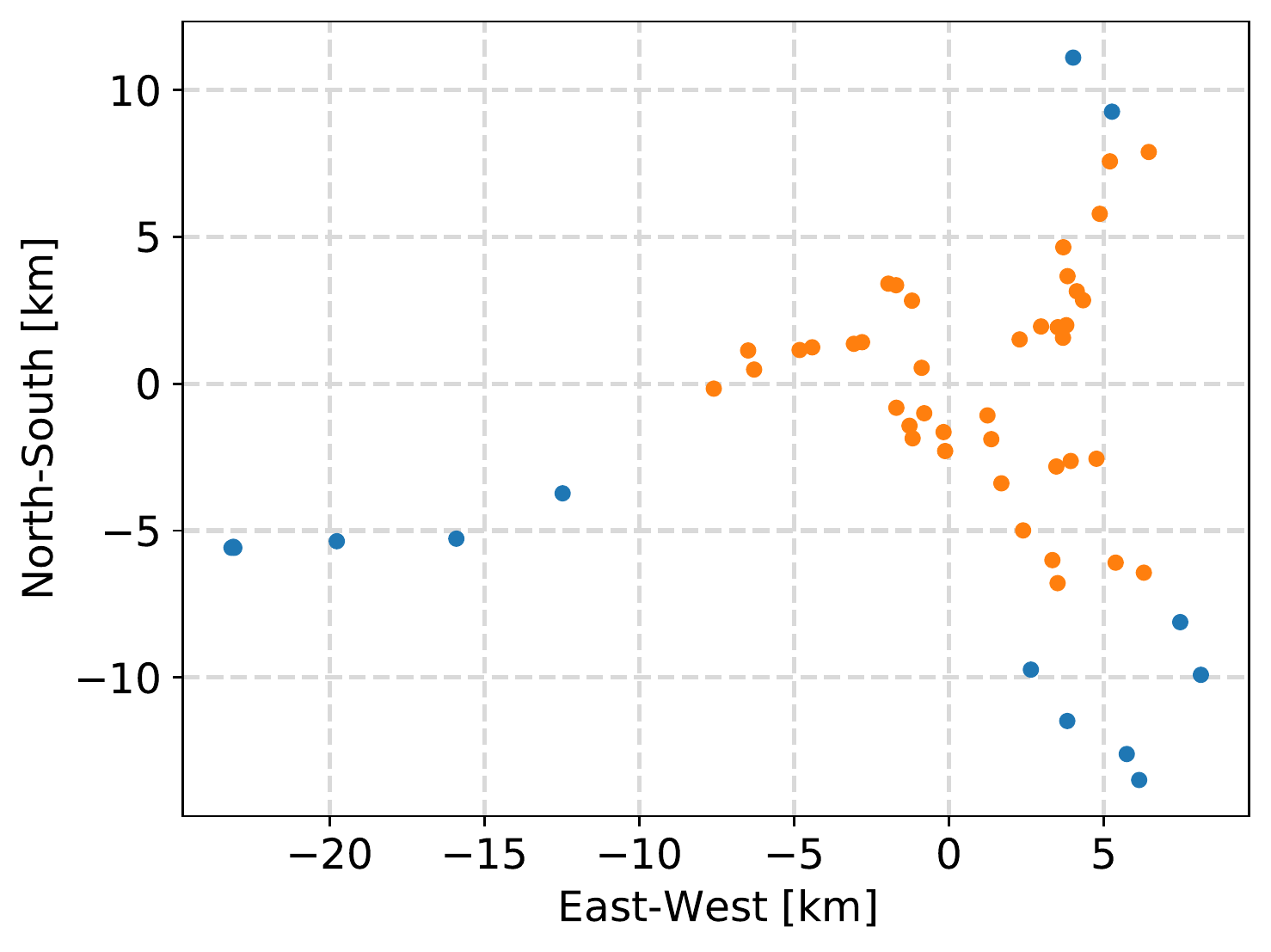}
    \end{center}
    \caption{\footnotesize{Plot of antenna positions for the extended ALMA configuration \texttt{ax54}.  Existing pads are shown in orange, new pads in blue.}}
    \label{fig:ax54_ants}
\end{figure}

\begin{figure}[hbtp]
    \begin{center}
    \includegraphics[width=\columnwidth]{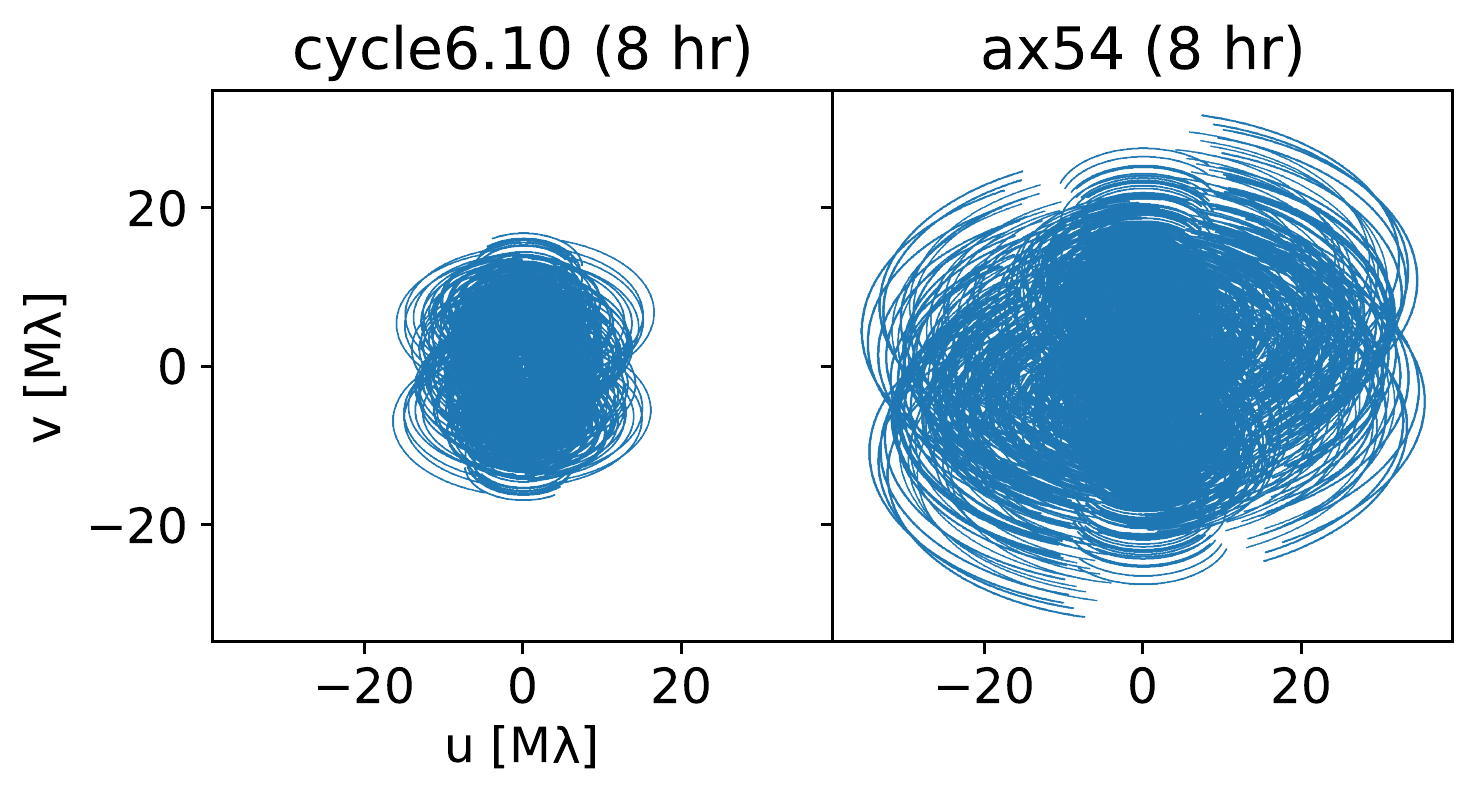}
    \end{center}
    \caption{\footnotesize{Comparison of uv-coverage for 8 hour aperture synthesis using the \texttt{cycle6.10} and \texttt{ax54} array configurations (left and right panels, respectively).}}
    \label{fig:uv_compare}
\end{figure}

To simulate observations with a plausible future extension to the current ALMA array, we produced a new configuration, called here \texttt{ax54}, which merges 42 existing antenna pads (orange dots in Figure~\ref{fig:ax54_ants}), as well as 12 possible new pads located within
the ALMA concession and the Atacama Astronomical Park that are being considered for additional baselines \citep[blue dots; see Figure~9 in the \textit{ALMA Development Program: Roadmap to 2030,}][]{Carpenter:2020}. All antennas have a diameter of 12 m.
Baseline lengths in this configuration
range from 44 m to 32.6 km.
We simulated aperture synthesis interferometric observations with an on-source time
of 8 hours
% on source times for compact config
for both the \texttt{ax54} and \texttt{cycle6.10} configurations. The uv-coverage obtained with these observations are shown in Figure~\ref{fig:uv_compare}.

% up to 15% flux loss
We complemented the observations with the previous two array configurations with a relatively short, 1-hour integration on a more compact configuration, namely \texttt{alma.cycle6.4},
with baselines ranging from 15 to 783 m, with the goal of minimizing the interferometric filtering out of the most extended scales of the disk emission.  Table~\ref{table:flux} quantifies the amount of flux loss due to interferometric filtering with the \texttt{alma.cycle6.10} and \texttt{ax54} configurations, \textit{without} the addition of \texttt{alma.cycle6.4}. 
With the addition of this more compact configuration, the amount of interferometric flux loss was typically reduced by a factor of $\approx 2$. 

\begin{table*}[h]
\centering
\begin{tabular}{ @{\extracolsep{3pt}} c c c c c c c c }
\hline\hline
    \multicolumn{4}{c}{\textbf{Model}} &
    \multicolumn{2}{c}{\textbf{cycle6.10}} &
    \multicolumn{2}{c}{\textbf{ax54}} \\
    \cline{1-4} \cline{5-6} \cline{7-8}
    
    ${\bf \Sigma_{g,0}}$ &
    ${\bf q}$ &
    ${\bf a_p}$ &
    \textbf{\textbf{Flux}} &
    \textbf{\textbf{Loss}} &
    \textbf{\textbf{RMS}} &
    \textbf{\textbf{Loss}} &
    \textbf{\textbf{RMS}} \\
    $[{\rm g/cm^2}]$ & $[M_{\oplus}/M_{\odot}]$ & [au] & [mJy] & & [$\mu{}$Jy/beam] & & [$\mu{}$Jy/beam]
    \\
    
%    $[\rm{g}/\rm{cm^2}]$ & $[M_{\oplus}/M_{\odot}]$ &
%    [au] &
%    [mJy] &
%    [\%] &
%    [$\mu$Jy/beam] &
%    [\%] &
%    [$\mu$Jy/beam] 
%    \\
    
    \hline
    
    300 & 1 & 2 & 69 &
    3.2\% & 11.2 &
    3.2\% & 8.4 \\
    300 & 10 & 2 & 67 &
    2.6\% & 9.7 &
    3.4\% & 8.4 \\
    300 & 1 & 3 & 97 &
    5.7\% & 9.9 &
    6.4\% & 8.3 \\
    300 & 10 & 3 & 94 & 
    5.6\% & 9.6 & 
    6.5\% & 8.3 \\
    300 & 1 & 5 & 142 &
    4.3\% & 9.5 &
    4.8\% & 8.4 \\
    300 & 10 & 5 & 137 &
    16\% & 10.8 &
    15\% & 8.7 \\
    100 & 1 & 2 & 42 &
    3.8\% & 9.7 &
    5.4\% & 8.4 \\
    100 & 10 & 5 & 80 &
    25.1\% & 10.5 &
    21.7\% & 8.4 \\
    2400 & 1 & 2 & 92 &
    2.2\% & 9.7 &
    2.4\% & 8.4 \\
    2400 & 10 & 5 & 186 &
    12.0\% & 10.8 &
    10.4\% & 8.5 \\
\hline\hline
\end{tabular}
\caption{\footnotesize{Total flux, flux loss with respect to the model flux and due to interferometric spatial filtering, and RMS noise for selected models, given an integration time of 8 hours and central wavelength of 0.88 mm.}}
\label{table:flux}
\end{table*}

\begin{table*}[h]
\centering
\begin{tabular}{ @{\extracolsep{3pt}} c c c c c c c c }
\hline\hline
    \multicolumn{4}{c}{\textbf{Model}} &
    \multicolumn{2}{c}{\textbf{cycle6.10}} &
    \multicolumn{2}{c}{\textbf{ax54}} \\
    \cline{1-4} \cline{5-6} \cline{7-8}
    
    ${\bf \Sigma_{g,0}}$ &
    ${\bf q}$ &
    ${\bf a_p}$ &
    \textbf{\textbf{Flux}} &
    \textbf{\textbf{Loss}} &
    \textbf{\textbf{RMS}} &
    \textbf{\textbf{Loss}} &
    \textbf{\textbf{RMS}} \\
    $[{\rm g/cm^2}]$ & $[M_{\oplus}/M_{\odot}]$ & [au] & [mJy] & & [$\mu{}$Jy/beam] & & [$\mu{}$Jy/beam]
    \\
    
    \hline

    300 & 1 & 2 & 30 &
    2.9\% & 6.6 &
    2.5\% & 5.0 \\
    300 & 10 & 2 & 29 &
    3.0\% & 6.6 &
    2.5\% & 5.1 \\
    300 & 1 & 3 & 43 &
    4.1\% & 6.3 &
    4.1\% & 5.4 \\
    300 & 10 & 3 & 41 &
    4.8\% & 6.8 &
    4.5\% & 5.5 \\
    300 & 1 & 5 & 66 &
    7.8\% & 6.6 &
    8.5\% & 5.5 \\
    300 & 10 & 5 & 62 &
    8.2\% & 6.6 &
    7.9\% & 5.3 \\
    100 & 1 & 2 & 17 &
    4.0\% & 6.3 &
    4.3\% & 5.1 \\
    100 & 10 & 5 & 33 &
    14.6\% & 6.6 &
    15.3\% & 5.6 \\
    2400 & 1 & 2 & 47 &
    2.8\% & 6.4 &
    2.5\% & 5.7 \\
    2400 & 10 & 5 & 100 &
    5.2\% & 6.6 &
    5.0\% & 5.3 \\
\hline\hline
\end{tabular}
\caption{\footnotesize{Same as Table~\ref{table:flux}, but for observations with a central wavelength of 1.25 mm.}}
\label{table:flux_1.25}
\end{table*}

% when flux is missing, will be specified when used

Simulated measurements of the interferometric visibilities
were carried out using the \texttt{simobserve} task
in \texttt{CASA}. We considered a total bandwidth of 7.5 GHz in a single spectral window (in the Appendix section we also investigate the effects of considering the spatial variation of the disk emission across 4 spectral windows, as in real ALMA observations, but we found only very modest differences on the final image). 
The simulated atmospheric noise parameters we used
correspond to good, i.e. under 2nd octile, weather conditions,
% https://library.nrao.edu/public/memos/alma/memo602.pdf
with an adopted value for the precipitable water vapour (PWV)
of 0.5 mm.
The ground temperature was set to 269 K.
%and the sky temperature to 260 K.

The \texttt{tclean} task
from \texttt{CASA}
was used to image the interferometric visibilities. 
We adopted a Briggs weighting scheme and after testing the imaging with different robust parameters $R$, we found that $R = -1$ provided the best results in terms of maximizing the signal-to-noise ratio of the disk substructures. 
% tried -2 as well, -1 was better.
% reproducing substructures at a good signal to noise ratio.
We also used a multiscale deconvolver
\citep[as in][]{Cornwell:2008}
with scales of 0, 7, and 30 pixels.
Since, as described in Section~\ref{sec:model}, the radial domain of our simulations depends on the planet orbital radius $a_p$ but the number of pixels in the maps is the same for all our models, then the pixel size in our  maps scales linearly with $a_p$ as:  $\rm{pixel~size} \rm{[\mu arcsec}] = 71.4$~$a_p$[au].

Table~\ref{table:ax54_res} lists the sizes of the synthesized beams obtained with a 8-hour aperture synthesis with the \texttt{ax54} configuration at 0.88 mm and at different source declinations. The chosen values are close to the mean declinations of young stars in nearby star forming regions, which are natural targets for studies of protoplanetary disks at high spatial resolution. Within the sample of regions considered here, which span declinations from $-77$\textdegree~to $+24$\textdegree, the beam major and minor axes range between $5.9-8.3$ mas ($0.9-1.2$ au) and $4.0-4.5$ ($0.6-0.7$ au) mas, respectively. The maps presented in this paper were derived assuming a declination of $-35$\textdegree.

\begin{table}[h]
\centering
\begin{tabular}{ c c c c c }
\hline\hline
    % multirow?
    % rearrange
    % add angle
    % change format to resemble others
    \textbf{Dec} &
    \textbf{Star-forming} &
    \multicolumn{3}{c}{\textbf{Beam}} \\
    \cline{3-5}
    & \textbf{region} & \textbf{MAJ} & \textbf{MIN} & \textbf{PA} \\
     $[$\textdegree$]$
     &  & [mas, au] & [mas, au] & [\textdegree] \\
\hline%\hline
    $+24$ & Taurus & 8.3, 1.2 & 4.3, 0.6 & $-5.1$ \\
\hline
    $-24$ & Ophiuchus & 6.6, 0.9 & 4.5, 0.6 & $-14$ \\
\hline
    $-35$ & Lupus & 6.3, 1.0 & 4.4, 0.7 & $-15$ \\
\hline
    $-77$ & Chamaeleon & 5.9, 0.9 & 4.0, 0.6 & $-8.1$ \\
\hline\hline
\end{tabular}
\caption{\footnotesize{FWHM beam sizes for an 8-hr integration using the \texttt{ax54} array configuration with Briggs robust weighting with $R = -1$ at a wavelength of 0.88 mm. The two numbers for the major and minor axes correspond to the angular (in mas) and spatial (in au) resolutions, respectively. The assumed distances are 140 pc for Taurus, 130 pc for Ophiuchus, 170 pc for Lupus, 160 pc for Chamaeleon.}}
\label{table:ax54_res}
\end{table}

We also compared the results
of the traditional \texttt{CLEAN} deconvolution algorithm (\texttt{tclean} task in \texttt{CASA})
to those produced by the
the \texttt{frankenstein} package
described in \citet{Jennings:2020}.
The \texttt{frankenstein} algorithm 
fits the real
visibilities as a
function of baseline distance
with a non-parametric Gaussian process.
In the case of an azimuthally symmetric source,
this function
is related to the radial profile of the source.
% Hankel transform
Consequently, this fit is able to
reconstruct radial profiles
directly from the visibilities
rather than producing them from a 2D image.
This avoids the loss of resolution, 
especially with long baselines,
from using the \texttt{CLEAN} algorithm
to subtract out sidelobes from the dirty image.
However, since \texttt{frankenstein} makes the assumption
of an azimuthally symmetric disk,
it is not valid for disks 
with significant azimuthal asymmetry
such as arcs or spirals.
Most, but not all, of our models only feature annular gaps,
lacking significant
azimuthal dependence in their substructures. In those cases,
\texttt{frankenstein} is
capable of retrieving information 
on their disk structure at sub-beam resolution. Although all the maps presented here were obtained using \texttt{tclean}, in Section~\ref{sec:results} we also show the comparison between the results with \texttt{tclean} and \texttt{frankenstein} obtained for one specific disk model without significant azimuthal asymmetries.

% make a table with declinations (+associated region of star form.)
% column angular res for both (at 1 hr and at 8 hr)
%     (maj and min ax of beam)
% columns for two different wavelengths
% don't bother if it's like 10-20% difference

% maybe add uv plots for ax54 and current alma
% custom antenna plots and uv plots

\section{Results}
\label{sec:results}

\begin{figure}[htbp]
    \begin{center}
    \includegraphics[width=\columnwidth]{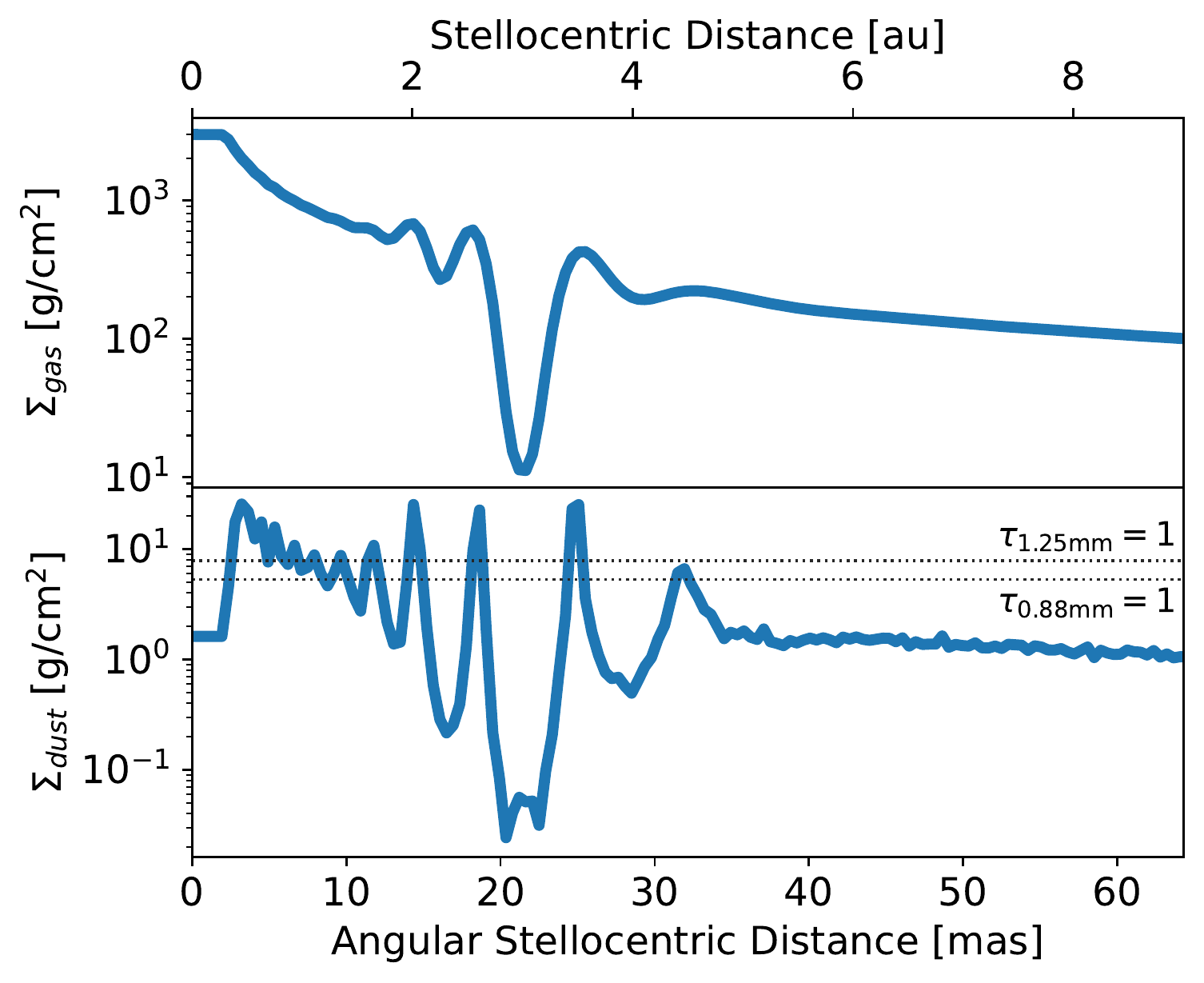}
    \end{center}
    \caption{\footnotesize{Radial profiles of the azimuthally averaged surface density for gas (top panel) and dust (bottom) for the model with $\Sigma_{g,0} = 300~\rm{g/cm^2}$, $q = 10~M_{\oplus}/M_{\odot}$ and $a_p = 3$~au. The two horizontal dotted lines in the bottom panel indicate the dust density values at which the dust optical depth $\tau_\lambda = 1$ at $\lambda$ = 0.88 and 1.25 mm, respectively.}}
    \label{fig:ref_model_gasdust}
\end{figure}

We will begin our discussion of the results
by examining the case of the disk model with
$\Sigma_{g,0}=300\,{\rm g/cm^2}$ and planet with $q=10\,{\rm M_{\oplus}/M_{\odot}}$
at an orbital radius of $3\,{\rm au}$.
Figure~\ref{fig:ref_model_gasdust} displays the radial profiles for the gas and dust surface densities of this disk model, after azimuthal averaging. The planet opens a pronounced gap in the gas around its orbital radius bracketed by two prominent density bumps but also less pronounced local density minima and maxima further from the planet. The local density maxima correspond to local pressure maxima, which are effective at slowing down the radial migration of grains and trapping the dust at those locations \citep[][]{Pinilla:2012}. As a consequence, the spatial distribution of dust density presents clear rings interleaved by regions of depleted dust (\textit{gaps}).   

\begin{figure*}[hbtp]
    \begin{center}
    \includegraphics{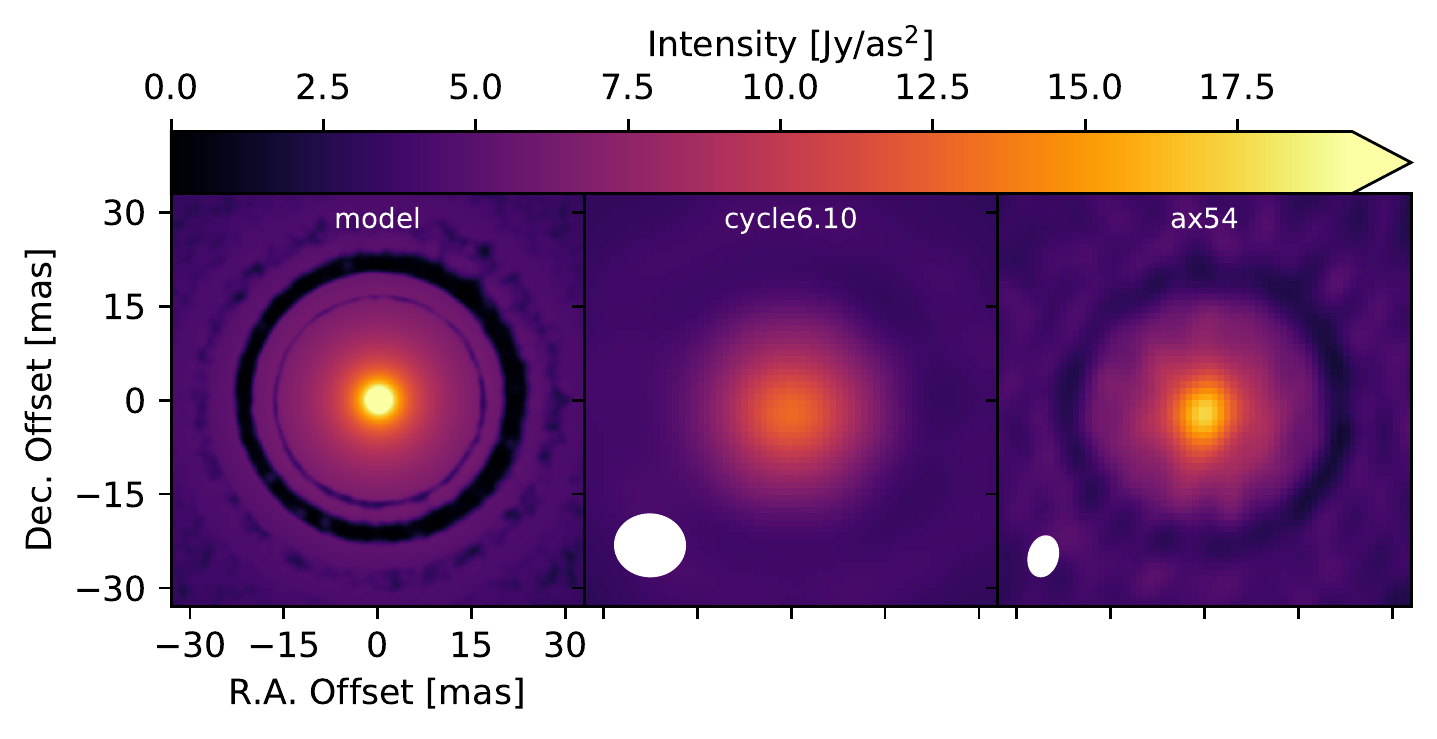}
    \end{center}
    \caption{\footnotesize{Maps for the dust continuum emission at 0.88 mm for the model with $\Sigma_0=300\,{\rm g/cm^2}$, $q=10~ M_{\oplus}/M_{\odot}$, $a_p=3\,{\rm AU}$. Left - Map of the disk model. Center - Simulated map with ALMA in \texttt{cycle6.10} configuration. The white ellipse in the bottom left corner shows the resolution beam, with sizes of 10.9 mas $\times$ 9.6 mas and position angle of $88$\textdegree. The rms noise on this map is 9.6~$\mu$Jy/beam.  Right - Simulated map with an extended ALMA in \texttt{ax54} configuration. The resolution beam in the lower left corner has sizes of $6.3$ mas $\times~4.4$ mas and position angle of $-15$\textdegree. The rms noise on this map is 8.3~$\mu$Jy/beam.}}
    \label{fig:ref_model_compare}
\end{figure*}

The dust continuum map extracted for this model at 0.88 mm, shown on the left panel in Figure~\ref{fig:ref_model_compare},
exhibits the primary gap at 3.0 au, corresponding to the orbital location of the planet,
flanked by the two secondary gaps at 2.3 and 3.9 au. No other gaps are visible because the emission from dust within the other less depleted gaps is optically thick. The central panel in the same figure shows the results of our ALMA simulations with 8 hours of integration time in the \texttt{cycle6.10} array configuration. The main gap produced by the planet is barely visible on the map, and at the resolution of these observations the gap is unresolved (beam size $\approx 3\times$ gap radial width). As a consequence, although the current ALMA array would detect this gap, no information on the planet properties could be extracted from the radial width and depth of the gap. 

The results displayed in the right panel show that the extended ALMA array is capable of spatially resolving the main gap, and also to detect the outer secondary gap at 3.9 au from the star. Because of the elongation of the synthesized beam, close to the North-South direction, this secondary gap is more clearly visible, although not spatially resolved, towards the East and West sides of the disk.

%\begin{figure}[hbtp]
%    \begin{center}
%    \includegraphics[width=\columnwidth]{pics/earth10_0.88mm_300gcm2_r-3_5000orb-close.pdf}
%    \end{center}
%    \caption{\footnotesize{Closeup of the dust continuum map at $\lambda =$ 0.88 mm for the disk model with $\Sigma_0=300\,{\rm g/cm^2}$, $q=10\,{\rm M_{\oplus}/M_{\odot}}$, $a_p=3\,{\rm AU}$.}}
%    \label{fig:model1}
%\end{figure}

\begin{figure*}[hbtp]
    \begin{center}
    \includegraphics{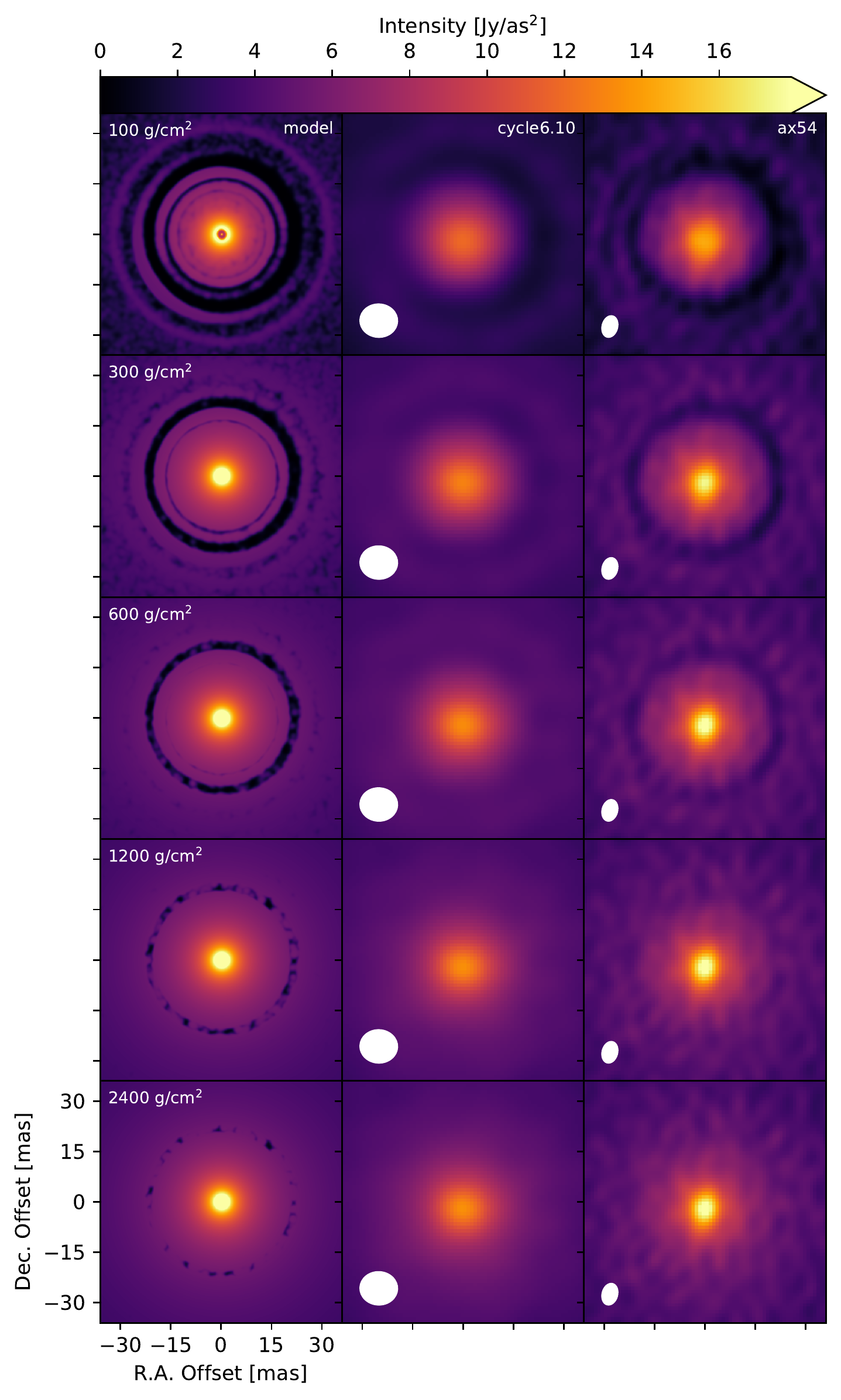}
    \end{center}
    \caption{\footnotesize{Same as in Figure~\ref{fig:ref_model_compare} but with models with different $\Sigma_0$ values as indicated on the left panel of each row. For all the models presented here, $q=10~ M_{\oplus}/M_{\odot}$, and $a_p=3\,{\rm au}$.}}
    \label{fig:sigma_grid}
\end{figure*}

Figure~\ref{fig:sigma_grid} shows the dust continuum maps for the models with the same $q$ and $a_p$ values as in Figure~\ref{fig:ref_model_compare} but with different $\Sigma_{g,0}$. The variation of this parameter has two main effects on the disk and its emission. Since the initial dust-to-gas ratio in our simulations is fixed, varying the gas densities leads to a change in the dust densities as well, with a consequent variation of the dust thermal emission. The fact that the surface brightness is relatively similar for the models presented in Figure~\ref{fig:sigma_grid} is due to the relatively high optical depth of the dust emission at 0.88 mm in the shown disk regions even for the models with lower $\Sigma_{g,0}$.  

The other main effect, which has a more significant impact on the maps derived from our models, is due to the dependence of the dynamics of dust on the local gas surface density. The Stokes parameter quantifies the aerodynamic coupling of dust with the gas, and a variation of this parameter can have a strong effect on the dynamics of the grains embedded in the disk. 
In particular, at a given grain size of the dust, which is fixed in our simulations, the Stokes parameter St $\propto 1/\Sigma_{g,0}$. Hence, the dust in disks with higher $\Sigma_{g,0}$ has lower St values. Since in all our models St $< 1$, this means that the dust in the denser disks is more dynamically coupled to the gas, hence the dust density and emission will appear radially smoother and more axisymmetric than in disks with lower densities. 

In fact, the only model showing clear azimuthally asymmetric substructures in Figure~\ref{fig:sigma_grid} is the one with lowest gas density ($\Sigma_{g,0} = 100~\rm{g/cm^2}$, top row), where the most prominent features are arcs instead of full rings as in the other models. The bright East side of the most prominent arc is detected with the extended ALMA, whereas the same feature appears to be blended with nearby dust emission in the current ALMA map, in which the only visible substructure is a wide asymmetric gap. The other rows in Fig.~\ref{fig:sigma_grid} show that the extended ALMA can detect the primary gap in the disks with $\Sigma_{g,0}$ values up to $\approx 1200~\rm{g/cm^2}$, while the gap is too narrow and with too little contrast to be detected in denser disks.         

\begin{figure*}[hbtp]
    \begin{center}
    \includegraphics{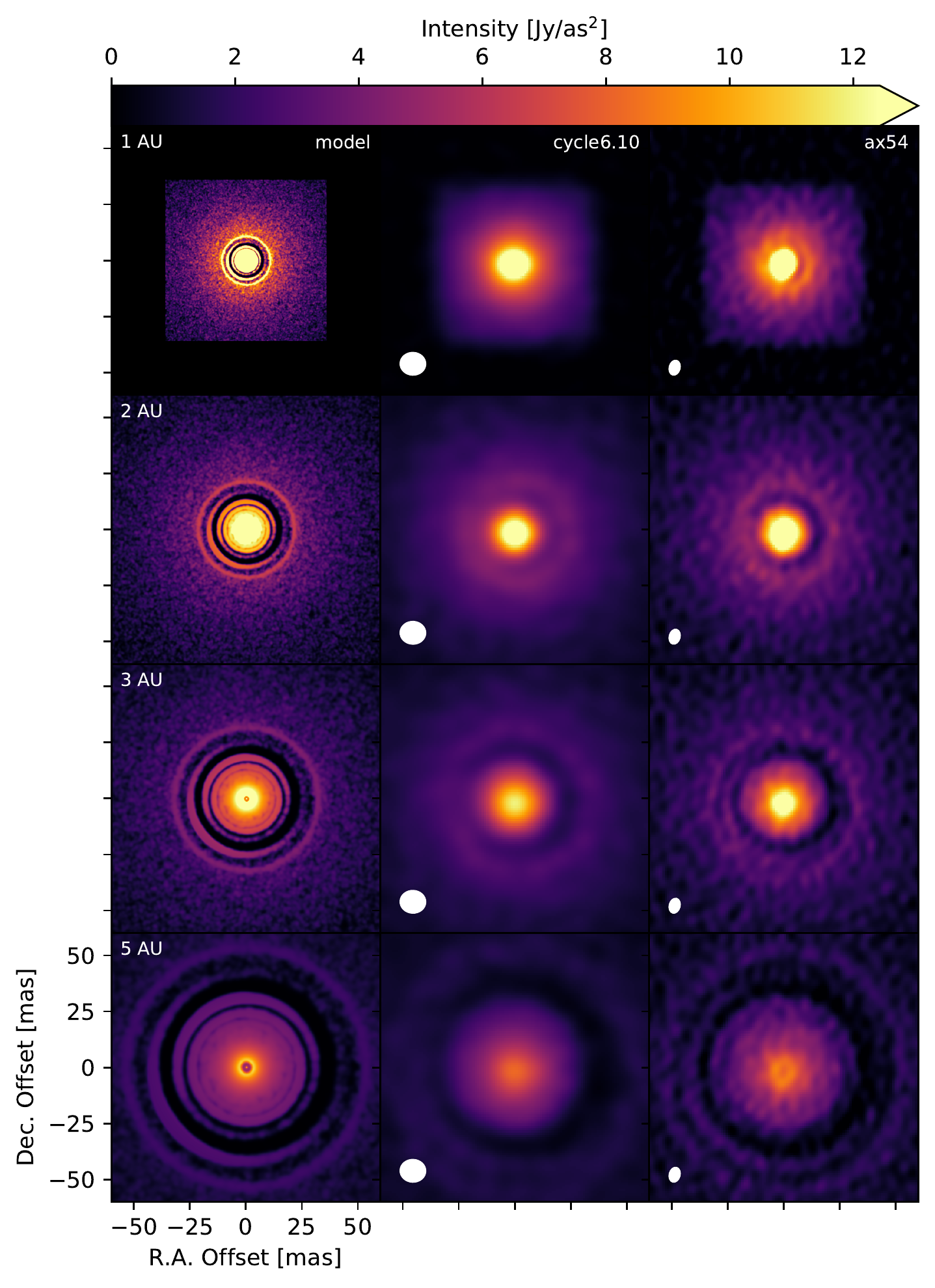}
    \end{center}
    \caption{\footnotesize{Same as in Figure~\ref{fig:ref_model_compare} but with models with different values for the orbital radius of the planet as indicated on the left panel of each row. For all the models presented here, $\Sigma_{g,0} = 100\,{\rm g/cm^2}$ and $q=10~ M_{\oplus}/M_{\odot}$.  The maps on the top row do not fill the entire size of the image because of the limited radial domain of our disk models, which scales with the orbital radius of the planet $a_p$, as described in Section~\ref{sec:methods}.}}
    \label{fig:radius_grid_earth10}
\end{figure*}

Figure~\ref{fig:radius_grid_earth10} displays the results for models with varying orbital radii of the planet, ranging from $a_p = 1$ au (top row) to 5 au (bottom row). The initial gas surface density at the planet orbital radius $\Sigma_{g,0}$ was set to 100~$\rm{g/cm^2}$ to maximize the number of disk substructures. For the model with $a_p = 5$ au, the extended ALMA detects and resolves multiple rings and asymmetric arcs, which the current ALMA does not resolve. The extended ALMA detects the gap also for the case of $a_p = 1$ au, even though the individual substructures are unresolved.  

\begin{figure*}[hbtp]
    \begin{center}
    \includegraphics{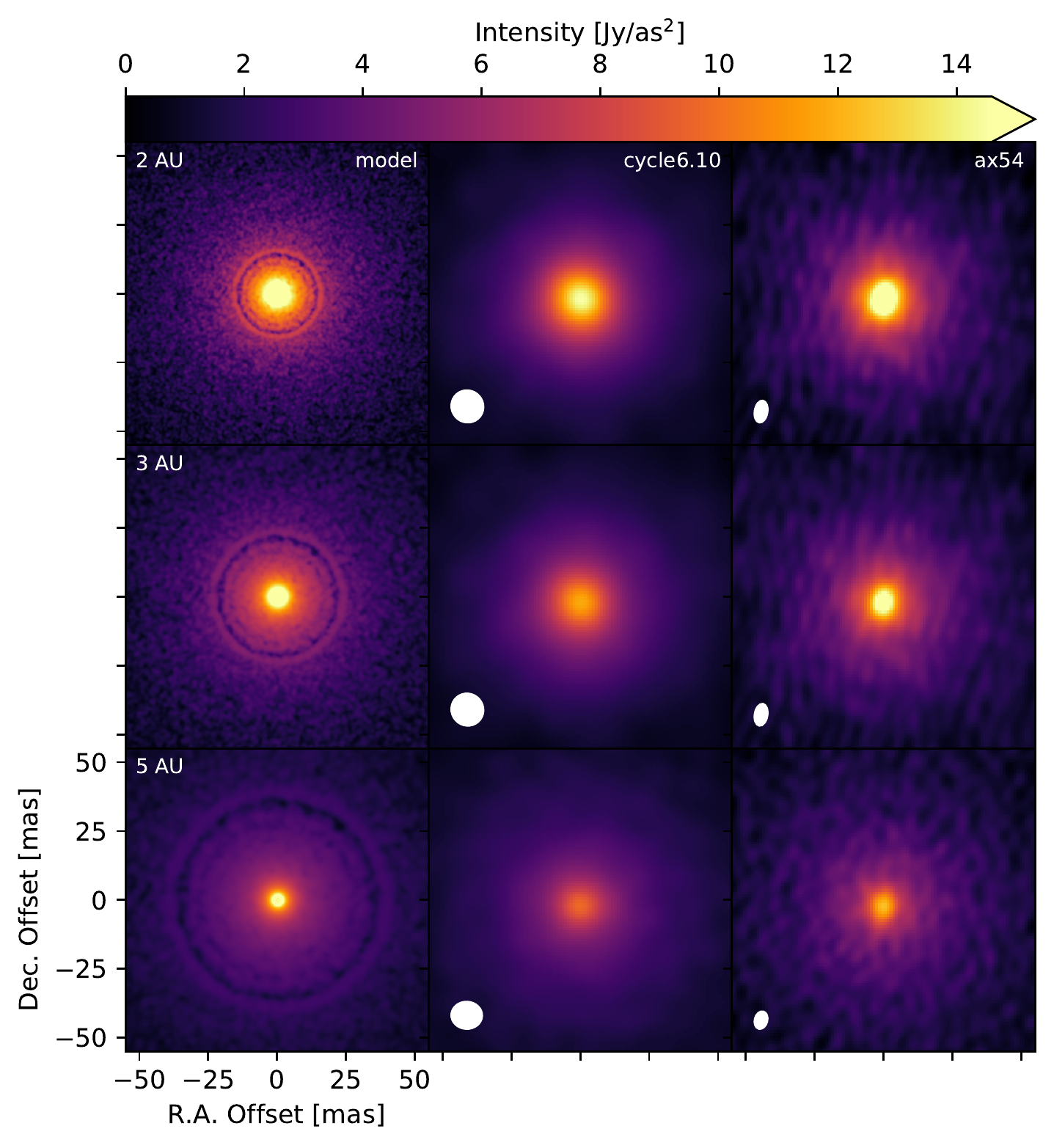}
    \end{center}
    \caption{\footnotesize{Same as in Figure~\ref{fig:ref_model_compare} but with models with different values for the orbital radius of the planet as indicated on the left panel of each row. For all the models presented here, $\Sigma_{g,0} = 100\,{\rm g/cm^2}$ and $q=1~M_{\oplus}/M_{\odot}$.}}
    \label{fig:radius_grid_earth1}
\end{figure*}

In Figure~\ref{fig:radius_grid_earth1} we present the models with $q = 1~M_{\oplus}/M_{\odot}$, $\Sigma_{g,0} = 100~\rm{g/cm^2}$ and for varying $a_p$. Compared with the maps shown previously, the predicted gaps are significantly shallower because of the lower gap radial width and depth expected for a lower mass planet. In our calculations, the gaps are not clearly detected with neither the current nor the extended ALMA, even though some level of depression in the surface brightness for the $a_p = 5$ au model is seen at low signal-to-noise ratio with the \texttt{ax54} configuration.

\begin{figure*}[hbtp]
    \centering
    \begin{tabular}{cc}
    \includegraphics[width=\columnwidth]{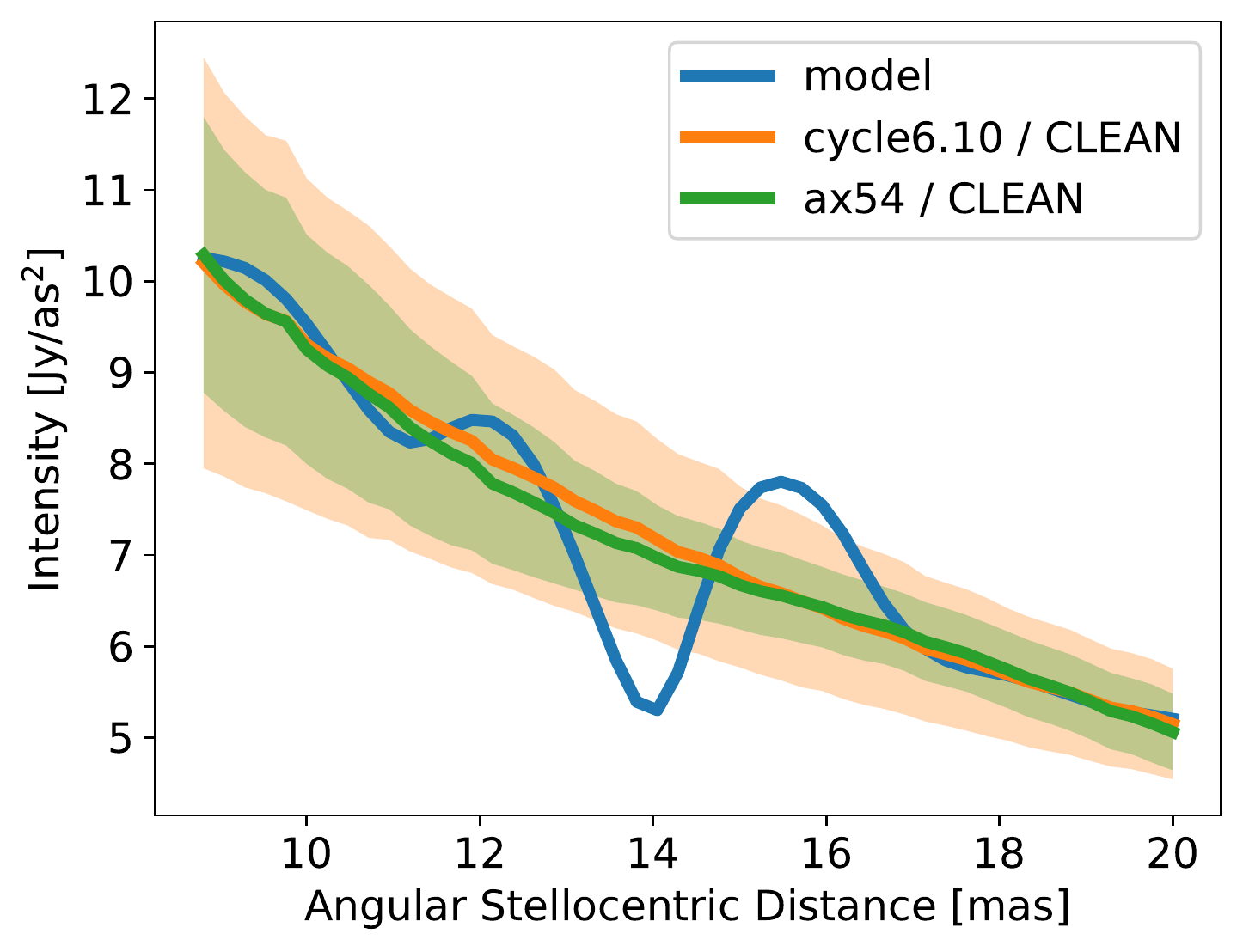} &
    \includegraphics[width=\columnwidth]{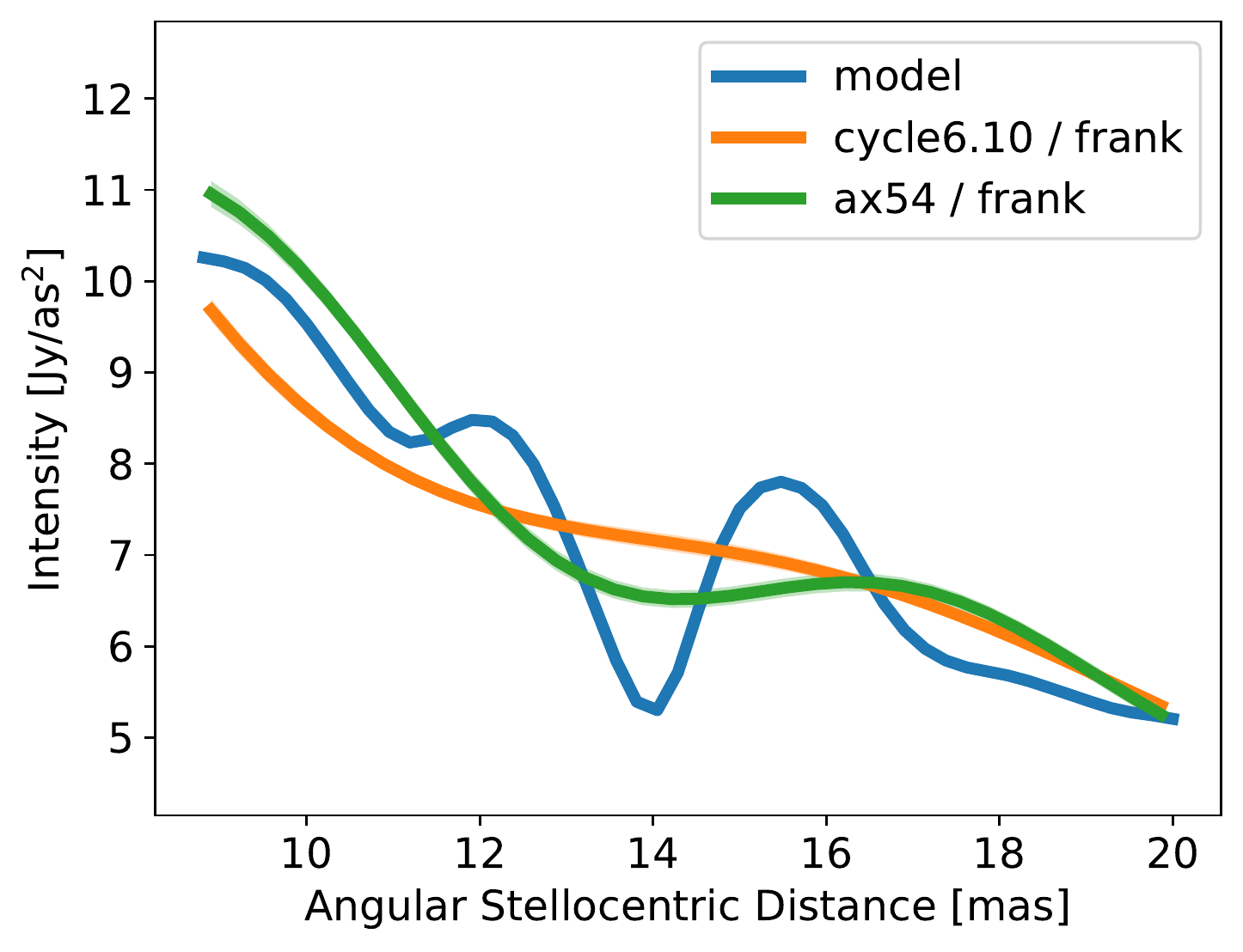}
    \end{tabular}
    \caption{\footnotesize{Radial profiles of the azimuthally averaged surface brightness at 0.88 mm for the model with $\Sigma_{g,0} = 100~\rm{g/cm^2}$, $q = 1~M_{\oplus}/M_{\odot}$, and $a_p = 2$~au. Blue line is for the model, orange and green lines are for the simulated observations with \texttt{cycle6.10} and \texttt{ax54} configurations, respectively. Left - Radial profiles extracted from the maps obtained with \texttt{tclean}. The error bars were obtained by dividing the standard deviation of the pixel values in each annular region by the square root of the number of synthesized beams fitting within the annular region. Right - Radial profiles and error bars obtained with \texttt{frankenstein}.}}
    \label{fig:low_mass_rp_frank}
\end{figure*}

When no significant azimuthal asymmetric features are present, \texttt{frankenstein} can be used to enhance the signal-to-noise of the annular rings and gaps that are otherwise not visible in the images obtained with deconvolution via \texttt{CLEAN}.
To demonstrate this, the left panel in Figure~\ref{fig:low_mass_rp_frank} shows the radial profile of the azimuthally averaged surface brightness produced by the \texttt{CLEAN} algorithm for a model with $\Sigma_{g,0} = 100~\rm{g/cm^2}$ and a $q = 1~M_{\oplus}/M_{\odot}$ planet at 2 au from the star.  No gap can be seen with either the current or extended configuration.
The right panel in the same figure shows the radial profiles obtained with \texttt{frankenstein}.  Although the uncertainties inferred by the \texttt{frankenstein} code are somewhat underestimated---%
a limitation described by \citet{Jennings:2020}---%
a gap is visually evident when analyzing the data from the extended configuration with \texttt{frankenstein}, which would otherwise be invisible using \texttt{CLEAN}.

\begin{figure*}[hbtp]
    \begin{center}
    \includegraphics{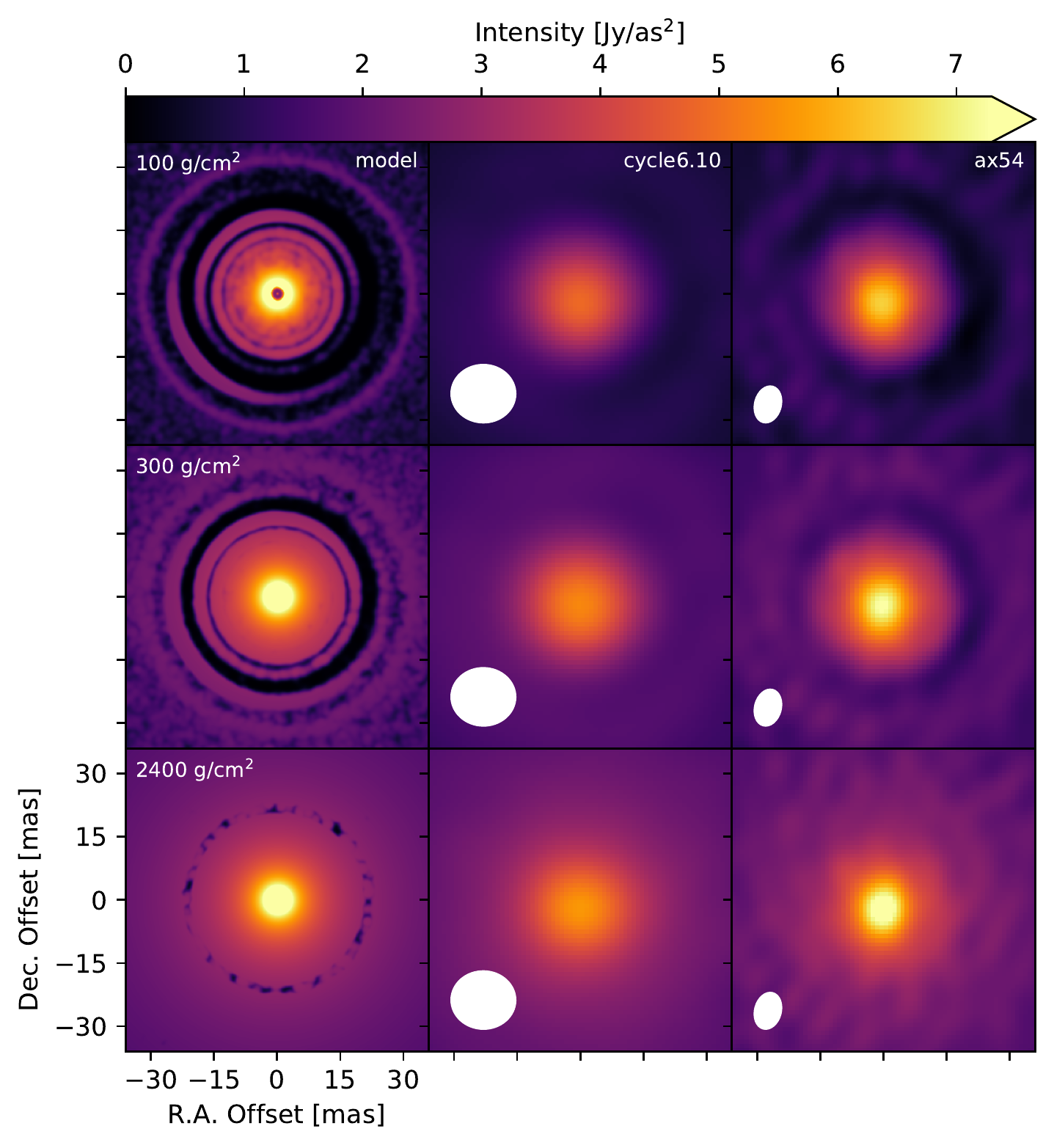}
    \end{center}
    \caption{\footnotesize{Similar to Figure~\ref{fig:sigma_grid} but for the dust continuum emission at a central wavelength of 1.25 mm. The sizes of the resolution beams are of 15.1 by 13.6 mas and 8.6 by 6.0 mas for the simulated observations with \texttt{cycle6.10} and \texttt{ax54} configurations, respectively.}}
    \label{fig:ref_model_compare_1.25}
\end{figure*}

All the maps discussed so far were derived at a wavelength of 0.88 mm. This wavelength is within the spectral coverage of ALMA Band 7, which for typical Spectral Energy Distributions of T Tauri stars provides the best sensitivity to the emission from dust in protoplanetary disks. Figure~\ref{fig:ref_model_compare_1.25} shows the dust continuum emission maps for models with varying $\Sigma_{g,0}$ values as in Figure~\ref{fig:ref_model_compare} but for a wavelength of 1.25 mm, which falls within the spectral coverage of ALMA Band 6. One benefit of observing disks at longer wavelengths is that their dust emission is more optically thin. Although with a lower signal-to-noise ratio than in Band 7, this figure shows that the gaps can be detected with the extended ALMA array also in Band 6, with a significant improvement over the similar observations with the \texttt{cycle6.10} configuration. In the Appendix section we show the expected improvements on the imaging of young disks when considering the capabilities of the new ALMA Band 6 receivers which are currently under development.
% \textbf{(We could replace Figure 6 with a multi-panel figure at 1.25mm to show multiple models in one figure, either with varying Sigma or the planet orbital radius.)}

\section{Discussion}
\label{sec:discussion}

The results presented in the previous Section show the potential of an upgraded ALMA with longer baselines by a factor of $2\times$ over the current ALMA capabilities to detect disk substructures due to planets at orbital radii of 5 au and lower in nearby star forming regions. For low values of the gas viscosity in these disk regions ($\alpha = 10^{-5}$), our simulations indicate that planets with $q = 1$ and $10~M_{\oplus}/M_{\star}$ are massive enough to open gaps that can be observed in the dust continuum emission with an upgraded ALMA.
We note here that for disks with higher gas viscosity, planets with higher masses than those investigated here would be needed to open gaps in the disk. Although this is an indirect method to probe planets embedded in the parental disk, it may be the best method, if not the only one, for several systems because of the expected high extinction of the planet light at optical/infrared wavelengths due to dust in the disk \citep[][]{Sanchis:2020,Asensio-Torres:2021}. 

The detection of young planets in these inner regions of protoplanetary disks would complement the results obtained so far with ALMA in the disk outer regions, i.e. at stellocentric radii larger than about 10 au \citep[see, for example, the results of the DSHARP Large Program,][]{Andrews:2018,Zhang:2018}. This would open the investigation to a more complete comparison between the properties of the detected exoplanet population and those of the young planets still embedded in the disk, which would shed light on the history of formation and migration of the known exoplanets~\citep[e.g.,][]{Zhang:2018,Lodato:2019,Kanagawa:2021}. In particular, the planets considered in this work are the likely precursors of the ice giant planets found at few au from their star, which \citet[][]{Suzuki:2016} claimed to be the most common types of planets from the results of microlensing surveys.

The disk model images presented in this work were obtained considering a distance of 140 pc, which is close to the average distance of young disks in several nearby star forming regions. 
In the Ophiuchus, Chamaeleon and Lupus star forming regions, 32, 14 and 15 disks have approximately the same total flux density (within $20\%$) or are brighter than our reference disk model at the wavelengths probed by the observations, i.e. 0.89 or 1.3 mm~\citep[][respectively]{Cieza:2019,Pascucci:2016,Ansdell:2016}. At the same time, most of these relatively bright disks are large, with outer radii in the dust emission larger than $\approx 50-100$ au and the possibility to detect disk substructures at similar SNR than those presented here will critically depend on the brightness of the dust emission within $\approx 5 - 10$ au from the star. Moreover, it is important to note that the majority of the disks detected in these observational surveys are not spatially resolved at resolutions of $30-50$ au, and it is possible that several of them are bright enough at stellocentric radii $< 10$ au to observe the planet-induced structures predicted by our models.

%\change{
%- how many disks could we see at this signal-to-noise ratios and higher; procedure:
%take a model (e.g. reference model) and derive its flux at 0.87mm or 1.3mm (same as observing wavelength), count the number of observed disks, divided by region, with a total flux comparable or higher than the disk model;
%then measure the flux from the model within the region with substructures (for example for the reference model that could be 4 or 5 au from the star), look at the unresolved observed disks and count how many of them have total flux comparable or higher than this disk model flux within 4-5au from the star; consider repeating this procedure for other disk models.
%}

The improved angular resolution of an extended ALMA would also reduce the minimum time interval required to detect proper motions of the disk substructures, e.g. the expected azimuthal motion of the extended arcs as well as of possible circumplanetary disks.

To use an example from our models, we consider the model with $\Sigma_{\rm{0}} = 100\,{\rm g/cm^2}$ shown in Figure~\ref{fig:sigma_grid}, which features a prominent asymmetric structure at an angular stellocentric radius $r =$ 25 mas (3.5 au), that is resolvable with the \texttt{ax54} configuration with ALMA Band 7.  The region close to the center of this arc can be well described by an azimuthal Gaussian function $I(\theta) = I_0 e^{-\theta^2 / 2 s^2}$ with parameters $I_0 = 4.2\,{\rm Jy / as^2}$, $s = 1.3\,{\rm rad}$ from the ALMA Band 7 image.
% Note: fitting parameters are very approximate here
At the peak intensity, the simulations presented here with the \texttt{ax54} configuration achieve a signal-to-noise ratio $I_0/\sigma \approx 15$.
The astrometric uncertainty in determining the peak of the Gaussian is then approximately equal to $\frac{r \times s}{I_0/\sigma} \approx 2.1\,{\rm mas}$. For a star with mass of $1\,M_{\odot}$, the required time interval to detect the proper motion of this structure at $5 \sigma$ is about 160 days.

In the case of a more compact point-like source such as the case of a circumplanetary disk detected at high signal-to-noise ratio (SNR $> 20$, corresponding to flux densities $> 170 {\rm~\mu{}Jy}$ in our ALMA Band 7 simulations), the astrometric precision is limited by atmospheric effects.  Under nominal observing conditions, the ALMA Technical Handbook\footnote{ \texttt{https://almascience.nrao.edu/documents-and-tools}}
states that the best astrometric uncertainty is given by $\sigma_{PSF}/20$, although since our beam size is also much finer than 0.15 arcsec, we must further increase this uncertainty by a factor of 2$\times$ due to the expected phase decorrelation. This results in an astrometric precision of $\approx 0.6$ mas in ALMA Band 7 with the \texttt{ax54} configuration. This gives a time interval of 44 days to detect the proper motion at $5\sigma$ for a bright point source at 3 au from a Solar-mass star at 140 pc, or 57 days for a point source at 5 au.

It is worth noting that although the only case of a clear circumplanetary disk known so far \citep[i.e. PDS 70c which is at an orbital radius of about 34 au,][]{Benisty:2021} would be detected with our ALMA Band 7 simulations with a SNR $\approx 10$, the SNR of the dust emission from circumplanetary disks in the inner disk may be expected to be significantly lower. This is because the size of circumplanetary disks is expected to be related to the radius of the Hill sphere which is proportional to the stellocentric radius. Since the dust emission of these disks is expected to be mostly optically thick, their flux at sub-mm/mm wavelengths would scale quadratically with the stellocentric radius.

The disk models considered in our work lack the resolution to include material orbiting the planet in a circumplanetary disk. According to the theoretical calculations presented in \citet[][]{Zhu:2018} for the expected flux densities of circumplanetary disks at sub-mm/mm wavelengths, our simulated ALMA Band 7 observations with the \texttt{ax54} configuration would have enough sensitivity to detect at 3$\sigma$ the dust emission from the disk of a Jupiter-mass planet with mass accretion rate $\dot{M}_p = 10^{-5}~M_{\rm{Jup}}/$yr and a disk radius larger than $\approx 0.05$ au at a distance of 140 pc, under the assumption that the disk heating is dominated by the gas viscosity with $\alpha$ values of either $10^{-2}$ or $10^{-3}$ (see their Figure 2). Under this assumption, the same ALMA Band 7 observations would detect a viscous heating dominated disk located at an orbital radius of 5 au from the star around a planet with $M_p\dot{M}_p > 10^{-7} M_{\rm{Jup}}^2/$yr and $M_p\dot{M}_p > 10^{-5.5} M_{\rm{Jup}}^2/$yr for disks with viscosity parameters $\alpha = 0.001$ and 0.01, respectively (see their Figure 4).

Although all the disk models presented here include planets interacting with the gas and dust in the disk, an extended ALMA would probe at sub-au resolution disk regions where different hydrodynamic instabilities have been proposed to drive the transfer of the disk angular momentum. In this context, the disk regions at stellocentric radii $1$ au $< r < 10$ au are particularly interesting as various purely hydrodynamic instabilities are predicted to be active, contrary to the case of Magneto-Rotational Instability \citep[e.g.,][]{Lyra:2019}. At least in some cases, it has been shown that these instabilities can significantly affect the dynamics of the dust, and create substructures that can be observed via high angular resolution observations in the dust continuum emission at sub-mm/mm wavelengths \citep[e.g.,][for the case of Vertical Shear Instability]{Flock:2020,Blanco:2021}. Moreover, some radial substructures can be expected if different mechanisms dominate at different radii across the disk as a consequence of their different efficiencies for the angular momentum transport \citep[][]{Pfeil:2019}.

\section{Conclusions}
\label{sec:conclusions}

We have evaluated the potential of an upgraded ALMA to observe at sub-au resolution the terrestrial planet forming regions of young disks in nearby star forming regions.  The upgraded ALMA considered here consists of an array configuration that combines 42 existing antenna pads with 12 new ones within
the ALMA concession and the Atacama Astronomical Park, and that would double the maximum baseline length ($b_{\rm{max}} \approx 32$ km).  To quantify the effects of this upgrade, we simulated observations of a family of disk models, perturbed by single planets with planet-to-star mass ratios of 1 and 10~${\rm M_{\oplus}/M_{\odot}}$ at orbital radii of $1-5$ au from the host star, located at a distance of 140 pc.

Our work shows that an extended ALMA would markedly improve upon the detection and investigation of the disk substructures in the dust continuum expected from the interaction between Earth-mass and Super Earths planets and the disk in regions of low gas viscosity. In particular, gaps associated to Super-Earth planets with $q = 10~M_{\oplus}/M_{\odot}$ can be detected as close as 1 au (7 mas) from the star, and azimuthally asymmetric structures can be resolved in disks with relatively low density. Evidence of gaps produced by Earth-mass planets with $q = 1~M_{\oplus}/M_{\odot}$ can be identified at stellocentric distances as short as $\approx 2-3$ au through an analysis of the interferometric visibilities, as shown here using the \texttt{frankenstein} code.  

An extended ALMA would also allow more accurate measurements of the main properties of gaps, e.g. radial widths and depletion factors, at radial distances down to $\approx 2-3$ au from the central star.  These observations would provide more accurate estimates of planet properties from the data-model comparison.  A broader parameter space of planets could be investigated in this way, which would better overlap with the distribution of of exoplanets observed around mature stars.

An ALMA upgrade would also be important for synergy with a future \textit{Next Generation Very Large Array} (ngVLA), which is expected to detect disk substructures at longer wavelengths than those presented here, i.e. 3 mm and longer, at resolutions of few milliarcsec~\citep[][]{Ricci:2018,Harter:2020,Blanco:2021}. The combination of ALMA and ngVLA would enable multiwavelength studies at high angular resolution which can be used to constrain the dust surface densities, temperatures and grain size distributions across the disk~\citep[e.g.,][]{Carrasco:2019}.

%Our work also demonstrates the utility of the \texttt{frankenstein} algorithm as a method for analyzing interferometric observations of protoplanetary disks.  Although the traditional CLEAN algorithm was not very effective at resolving structures from any of our $q = 1~{\rm M_{\oplus}/M_{\odot}}$ models, using either current or extended ALMA, the high azimuthial symmetry of the gaps associated with these low-mass planets in our models made \texttt{frankenstein} an effective method to resolve such gaps.

%The capabilities of the proposed ALMA upgrade for the purpose of resolving substructures in the dust continuum emission of protoplanetary disks would be lower than those of more ambitious future projects such as a next generation Very Large Array, which would have even better angular resolution.  However, in regions of the sky covered by both observatories, an extended ALMA would still greatly complement ngVLA when or if both projects are realized because ALMA observes at smaller wavelengths than ngVLA would.  Since current ALMA comes no where close to matching the angular resolution of ngVLA, an extended ALMA would be needed to observe in the submillimeter some of the structures ngVLA would be able to detect in radio.

% 

% \vskip 0.1in

\acknowledgements We want to thank Crystal Brogan and John Carpenter for useful discussions on this project. We thank Neil Phillips and Rüdiger Kneissl for providing us with the antenna position file for the extended ALMA configuration. L. R. and Z. Z. acknowledge financial support from the ALMA North-American Cycle 7 Development Studies. ALMA is a partnership of ESO (representing its member states), NSF (USA) and NINS (Japan), together with NRC
(Canada), MOST and ASIAA (Taiwan), and KASI (Republic of Korea), in cooperation with the Republic
of Chile. The Joint ALMA Observatory is operated by ESO, AUI/NRAO and NAOJ. The National Radio Astronomy Observatory is a facility of the National Science Foundation operated
under cooperative agreement by Associated Universities, Inc.

\software{Astropy \citep[][]{Astropy:2013}, CASA \citep[][]{McMullin:2007}, frankenstein \citep[][]{Jennings:2020}, and the NASA Astrophysics Data System (ADS).}

\bibliographystyle{aasjournal}
\begin{singlespace}
\bibliography{main}
\end{singlespace}

% citations:
%\citep[][]{Zhang:2018}
%\citep[pre][post]{Zhang:2018}
%\citet{Zhang:2018}

\appendix
\label{sec:appendix}

We investigate here the improvement expected for the dust continuum imaging of one of our disk models when considering the capabilities of the new ALMA Band 6 receivers, which are currently under development\footnote{\texttt{https://science.nrao.edu/facilities/alma/alma-develop-old-022217/2nd\%20Gen\%20Band\%206\%20Rcvr}}~\citep[][]{Asayama:2020}. These receivers will improve the sensitivity of the current ones by a factor of $\sim \sqrt{2} \approx 1.4$, and will also provide a wider bandwidth by a factor of $\sim 2$. The extended bandwidth will in turn result in an extra improvement on the continuum sensitivity of $\sqrt{2}\times$, and will also provide better imaging fidelity thanks to the additional sampling of the interferometric $(u,v)$ space by spreading out the frequency coverage.

\begin{figure*}[hbtp]
    \begin{center}
    \includegraphics[scale=1.3]{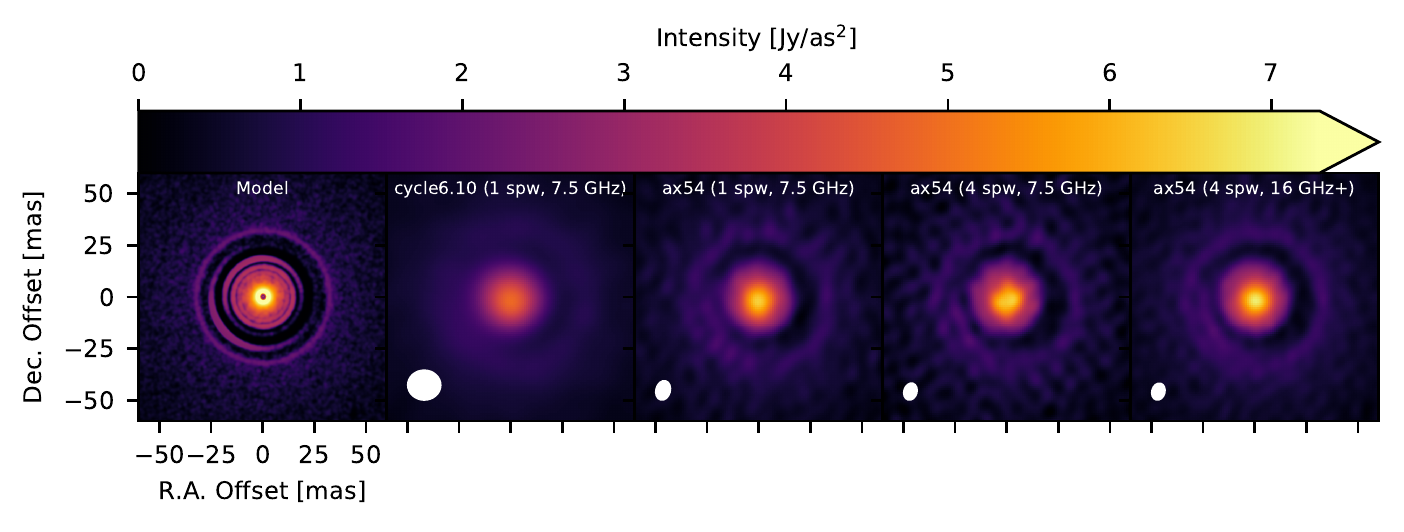}
    \end{center}
    \caption{\footnotesize{Maps for the dust continuum emission at 1.25 mm for the model with $\Sigma_0=100\,{\rm g/cm^2}$, $q=10~ M_{\oplus}/M_{\odot}$, $a_p=3\,{\rm au}$. The first three maps from the left are the same as in the top row of Figure~\ref{fig:ref_model_compare_1.25}. The fourth and fifth maps were obtained considering the dust emission in four spectral windows as described in the text. The simulations for the fifth map consider a total bandwidth of 16 GHz as well as a factor of $\sqrt{2}\times$ improvement in the sensitivity of the new ALMA Band 6 receivers.
    The size of the synthesized beam for both of the simulated \texttt{ax54} observations with 4 spectral windows is $7.4$ mas $\times~5.4$ mas.
    The rms noise values for the observed maps are 6.3, 5.4, 6.5 and 3.3 $\mu$Jy/beam, from left to right, respectively.}}
    \label{fig:sens_upgrade_close}
\end{figure*}

To simulate the results of future ALMA observations with the extended \texttt{ax54} configuration and with the new ALMA Band 6 receivers we adopted the following procedure. We considered the disk model with $\Sigma_{\rm{0}} = 100~$g/cm$^2$, $q = 10~ M_{\oplus}/M_{\odot}$, $a_p = 3$ au at a central frequency of 240 GHz (top row in Figure~\ref{fig:ref_model_compare_1.25}, with maps that are shown also in the left three maps in Figure~\ref{fig:sens_upgrade_close}). Whereas all the simulated observations presented in our work considered a total bandwidth of 7.5 GHz within a \textit{single} spectral window, here we simulate observations in 4 spectral windows centered at 231, 233, 247 and 249 GHz, respectively, with bandwidth for each window of 1.875 GHz, for a total bandwidth of 7.5 GHz. For each central frequency of each window we calculated an image from our disk model, using the same method as described in Section~\ref{sec:model}, in order to take into account the variation with frequency of the dust emission from the disk model. These images were then combined in our ALMA simulations via the \textit{Multi-Term Multi-Frequency} method with \texttt{nterms=2} in \texttt{tclean}.
Other than providing a more accurate representation of the dust emission across the bandwidth sampled by the observations, this procedure has the main advantage of obtaining better image fidelity thanks to the additional coverage of the $(u,v)$ space. The result of this simulation is shown in the fourth map from the left in Figure~\ref{fig:sens_upgrade_close}. The comparison with the third map in the same figure shows the modest changes with respect to the image obtained in Section~\ref{sec:results} which considered a single spectral window, and no variation with frequency of the disk emission.

A significant improvement is obtained when the same procedure was used to simulate the observations with the upgraded Band 6 receivers. To account for the $2\times$ wider bandwidth, we considered 4 spectral windows centered at 230, 234, 250 and 254 GHz, respectively, and each with a bandwidth of 4 GHz (total bandwidth of 16 GHz).  
In order to reproduce the additional improvement to the receiver sensitivity of $1.4\times$ without introducing any extra $(u,v)$ coverage, we simulated the same observation twice, with different \texttt{seed} parameters in \texttt{simobserve} to get different noise characteristics for the two simulations. We then concatenated the visibility datasets from these two simulations and imaged the interferometric visibilities with the \textit{Multi-Term Multi-Frequency} method in \texttt{tclean} as described above. The rightmost map in Figure~\ref{fig:sens_upgrade_close} displays the clear improvement in signal-to-noise ratio and image fidelity of the most prominent disk substructures predicted by the model.

% Beam sizes (mas x mas): 15.1 x 13.6, 8.6 x 6.0, 7.4 x 5.4, 7.4 x 5.4
% 1 spw, 7.5 (cycle6.10): 6.3 uJy/beam
% 1 spw, 7.5 (ax54): 5.4 uJy/beam
% 4 spw, 7.5: 6.5 uJy/beam
% 4 spw, 16: 3.3 uJy/beam

%\begin{figure*}[hbtp]
%    \begin{center}
%    \includegraphics{pics/sens-upgrade.compare.pdf}
%    \end{center}
%    \caption{\footnotesize{Full}}
%    \label{fig:sens_upgrade_full}
%\end{figure*}

\end{document}